\documentclass[a4paper,fleqn]{cas-dc}
\usepackage{fancyhdr}
\fancypagestyle{plain}{
  \fancyhf{} 
}
\usepackage{hyperref}

\usepackage{amsmath,amsfonts}
\usepackage{graphicx} 
\usepackage{stfloats} 
\usepackage{color} 
\usepackage{caption} 
\usepackage{algorithmic}
\usepackage{array}
\usepackage{pifont}
\usepackage{appendix}
\usepackage[caption=false,font=normalsize,labelfont=sf,textfont=sf]{subfig}
\usepackage{textcomp}
\usepackage{colortbl}
\usepackage{xcolor}
\usepackage{booktabs}
\usepackage{url}
\usepackage{hhline}
\usepackage{multirow}
\usepackage{makecell}
\usepackage{float}
\usepackage{verbatim}
\usepackage{vcell}
\usepackage{cite}
\usepackage{color,soul}
\usepackage{graphicx}
\usepackage{longtable}
\usepackage{tabularray}
\usepackage{xcolor}
\usepackage{atbegshi}   

\AtBeginShipoutNext{\AtBeginShipoutDiscard}   
\makeatletter
\def\ps@plain{%
  \let\@mkboth\@gobbletwo
  \def\@oddhead{\hfil\thepage} 
  \def\@evenhead{\thepage\hfil} 
  \let\@oddfoot\@empty 
  \let\@evenfoot\@empty 
}
\makeatother
\usepackage[numbers]{natbib}
\usepackage{balance}

\def\tsc#1{\csdef{#1}{\textsc{\lowercase{#1}}\xspace}}
\tsc{WGM}
\tsc{QE}

\makeatletter
\renewcommand\subsection{\@startsection{subsection}{2}{\z@}%
   {-3.25ex\@plus -1ex \@minus -.2ex}%
   {1.5ex \@plus .2ex}%
   {\normalfont\large}} 
\renewcommand\subsubsection{\@startsection{subsubsection}{3}{\z@}%
   {-3.25ex\@plus -1ex \@minus -.2ex}%
   {1.5ex \@plus .2ex}%
   {\normalfont\normalsize}} 
\makeatother
\begin{document}
\let\WriteBookmarks\relax
\def\floatpagepagefraction{1}
\def\textpagefraction{.001}
\renewcommand{\figurename}{Fig.}\


\title [mode = title]{Evaluation of LLM Chatbots for OSINT-based Cyber Threat Awareness}  



%



\ead{E-mail addresses: sshafee@ciencias.ulisboa.pt (S. Shafee), anbessani@ciencias.ulisboa.pt (A. Bessani),
pmf@ciencias.ulisboa.pt (P.M. Ferreira).}



\author{\textcolor{black}{Samaneh Shafee\textsuperscript{\hyperlink{corrauth}{*}}, Alysson Bessani, Pedro M. Ferreira}}



\affiliation[]{organization={LASIGE, Faculdade de Ciências, Universidade de Lisboa},
            country={Portugal}}

\cortext[1]{Corresponding author at: LASIGE, Faculdade de Ciências, Universidade de
Lisboa, Portugal.}
\hypertarget{corrauth}{\emph{Corresponding author: Samaneh Shafee - email@example.com}}

\begin{abstract}
Knowledge sharing about emerging threats is crucial in the rapidly advancing field of cybersecurity and forms the foundation of Cyber Threat Intelligence (CTI).
In this context, Large Language Models are becoming increasingly significant in the field of cybersecurity, presenting a wide range of opportunities.
This study surveys the performance of ChatGPT, GPT4all, Dolly, Stanford Alpaca, Alpaca-LoRA, Falcon, and Vicuna chatbots in binary classification and Named Entity Recognition (NER) tasks performed using Open Source INTelligence (OSINT).
We utilize well-established data collected in previous research from Twitter to assess the competitiveness of these chatbots when compared to specialized models trained for those tasks.
In binary classification experiments, Chatbot GPT-4 as a commercial model achieved an acceptable $\text{F}_1$ score of 0.94, and the open-source GPT4all model achieved an $\text{F}_1$ score of 0.90. However, concerning cybersecurity entity recognition, all evaluated chatbots have limitations and are less effective. 
This study demonstrates the capability of chatbots for OSINT binary classification and shows that they require further improvement in NER to effectively replace specially trained models.
Our results shed light on the limitations of the LLM chatbots when compared to specialized models, and can help researchers improve chatbots technology with the objective to reduce the required effort to integrate machine learning in OSINT-based CTI tools.
\end{abstract}



\maketitle
\setcounter{page}{1}
\section{Introduction}
\label{Introduction}
Cybersecurity, a continuously evolving domain, involves experts publicly sharing their knowledge on cyber threats.
This information is the primary source for Cyber Threat Intelligence (CTI) tools, and researchers have contributed to developing methods for extracting cyber threat intelligence from text sources (e.g.,~\cite{liao2016acing,ritter2015weakly}).
With the development of Machine Learning (ML), significant advancements have been made in the field of cybersecurity.
ML-based tools have deepened our understanding of cyber threats, enabling innovative and responsive solutions.
A notable development in this domain is the emergence of Large Language Models (LLM), representing a substantial breakthrough in Natural Language Processing (NLP).
This development has empowered chatbots to extract meaningful information about cybersecurity threats.
Such advancements are exemplified by academic works such as SecBot~\cite{franco2020secbot} and \citeauthor{arora2023developing}~\cite{arora2023developing}, which provide proof-of-concept chatbots to support cybersecurity planning and management, and recent high-profile products such as Microsoft Security Copilot~\cite{microsoft_seccopilot}.
These efforts (among others, described in Section~\ref{RWgeneral}) highlight the practical use of chatbots to address cybersecurity and CTI challenges.

In today's cybersecurity landscape, the timely detection and response to emerging threats are crucial.
To address this vital requirement, we focus on researching the potential of LLM-based chatbots to improve cyber threat awareness and streamline the detection processes.
Specifically, we investigate using pre-trained LLM chabots for binary classification and Named Entity Recognition (NER) tasks within the scope of generating indicators of compromise for CTI.
This integration enhances the field's capabilities, enabling organizations to strengthen their defences and proactively mitigate potential risks.
The important question remains of whether LLM-based chatbots can achieve the level of performance~\cite{KOCON2023101861} of specialized models~\cite{10.1145/3439726}, which are known to achieve excellent results using a wide variety of techniques for binary classification~\cite{li2022survey}, and NER~\cite{jehangir2023survey} tasks.

%
%
%
%
This paper presents an empirical study that uses inductive reasoning and a comparative methodology to answer the following research question: \emph{Are LLM chatbots competitive with state-of-the-art specialized models for detecting OSINT CTI and extracting pertinent information?
To produce the empirical results and inductively assess the performance of chatbots, we use a publicly available annotated dataset \cite{alves2021processing,dionisio2019cyberthreat} collected from Twitter, a known reliable OSINT CTI source \cite{alves2020follow}.
This dataset provides an accurate testing environment for challenging chatbots in a real-world CTI scenario.
Then, a comparative analysis of previous work and this paper's results effectively answers the research question, allowing the discussion of the conclusion's implications and outlining possibilities for further research and improvements.}

The LLM-based chatbots in this study include ChatGPT \cite{noauthor_openai_nodate}, GPT4all \cite{anand2023gpt4all}, Dolly \cite{Dolly}, Stanford Alpaca \cite{taori2023alpaca}, Alpaca-LoRA \cite{alpaca-lora}, Falcon \cite{noauthor_falcon_nodate}, and Vicuna \cite{chiangvicuna}.
These systems represent the two prominent types of chatbots widely used nowadays: paid chatbots available as a service through APIs, and Open-source chatbot models that are built to run on local GPU servers.
When planning the experiments, we picked all the available variants of GPT-like models since they are standard in the field~\cite{LOPEZESPEJEL2023100032}.

Since utilizing LLM chatbots raises additional technical questions, we also study the impact of different chatbot utilization methods on the empirical results obtained to answer the main research question.
These questions concern the capability of chatbots to provide clear yes-or-no answers, and the cost (in terms of processing time) required for these chatbots to perform the NLP tasks.

Our contributions can be summarized as follows:
\begin{enumerate}    
\item {We present a state-of-the-art survey on chatbot models and their application to cybersecurity}.
\item We investigate the extent to which the inherent flexibility of LLM-based chatbots can be tailored to meet the specific requirements of OSINT-based CTI applications.
\item {The study provides a comparative analysis of the practical use and performance of chatbots in specialized CTI tasks, including binary text classification and NER.}
\end{enumerate}

The remainder of this paper is organized as follows. 
Section \ref{RWgeneral} provides some background on LLMs and chatbots, and reviews related works on LLM chatbots for cybersecurity, and the evaluation of chatbots in NLP tasks. 
Section \ref{LLM-based Chatbots} presents a deep exploration of LLM-based chatbots and highlights their significance and capabilities.
In Section \ref{Chatbot Evaluation}, we shift our focus to the methodology used to evaluate chatbots, detailing our dataset, methods, and the comparison criteria employed for evaluation. 
Section \ref{Strategies for Optimal Utilization} explores strategies aimed at optimizing the utilization of these chatbots, including prompt fine-tuning and text length control.
Section~\ref{Experimental Results} presents our main experimental results and their discussion.
Finally, we conclude our study in Section \ref{Conclusion} by summarizing the key contributions and insights derived from this study.

\section{Background and Related work}
\label{RWgeneral}

In this section, we briefly present the required background for this paper.
We start by analyzing transformer models, the foundational elements facilitating recent progress in NLP.  
Next, we discuss LLMs by examining their development, abilities, and significant influence on various specific areas.
Building on this foundation, we review research on how LLMs can effectively be utilized for cybersecurity-related NLP, exploring in detail the role of OSINT in cybersecurity. 
We end the section by examining chatbots' evolution and NLP capabilities to address cybersecurity concerns and the literature gap we aim to contribute with this paper.

\subsection{Transformers}
The introduction of transformers~\cite{vaswani2017attention} revolutionized the field of NLP, as they became the preferred architecture for various NLP tasks due to their ability to effectively capture the extensive dependencies and contextual associations of textual data \cite{choi2023transformer,lin2022survey}.
This ability enables transformers to overcome the limitations of previous methods such as Recurrent Neural Networks (RNNs) \cite{medsker2001recurrent}, Convolutional Neural Networks (CNNs) \cite{kim2014convolutional}, and Long Short-Term Memory (LSTM) \cite{hochreiter1997long}.
Instead of processing inputs one at a time, transformer models handle tokens or words simultaneously.
This makes it easier to model the global interactions \cite{farooq2021global,sanford2024representational} and dependencies \cite{lin2022survey}.
This feature makes transformers highly effective for various tasks, including but not limited to machine translation, sentiment analysis, and text generation.
Transformers have emerged as the basic foundation for models such as Bidirectional Encoder Representations Transformers (BERT) \cite{devlin2018bert} and Generative Pre-trained Transformers (GPT) \cite{radford2018improving}.
These models have demonstrated exceptional performances across diverse NLP benchmarks, leading to progress in language understanding, text generation, and question answering \cite{min2023recent}.

\subsection{Large Language Models}

LLMs have emerged as significant innovations that revolutionize the human language process, generation, and understanding.
These models are trained on large text datasets, including immense volumes of language data, which enables them to perform tasks that demand contextual comprehension and generate coherent and meaningful responses \cite{10.1145/3649506}.
In terms of applications, the latest generation of language models has been applied across diverse fields.
For instance, the utilization of LLMs within mathematics, physics, and chemistry problem-solving has been evaluated by \citeauthor{arora2023have} \cite{arora2023have}.
\citeauthor{agrawal2023llms} evaluated the ability of LLMs for human-like reasoning tasks \cite{agrawal2023llms}. 
The findings of the study indicate that LLMs have strong abilities in analogical and moral reasoning but face challenges in spatial reasoning tests.
Chatbots powered by LLMs have attracted interest as powerful tools for data annotation in NLP domains \cite{ding2022gpt}.
This interest arises from chatbots' proficiency in language tasks and the critical role of data annotation in developing NLP systems.
An illustrative study \cite{gilardi2023chatgpt} compared ChatGPT with human crowd workers in annotation tasks.
The research emphasized that ChatGPT surpassed human workers in terms of performance, agreement among annotators, and cost-effectiveness.

There are two widely used types of LLMs: GPT and BERT.
GPT-style models utilize a regressive transformer architecture to capture contextual dependencies and relationships within text~\cite{10.1145/3649506}, while BERT-style language models utilize a bidirectional transformer architecture and adopt a masked language modelling objective.
Both approaches gained popularity for NLP capabilities because of their ability to model contextual dependencies.
Besides the difference in the context they consider to make text predictions, they also differ in the number of parameters and training dataset size, with GPT-style models requiring more resources in both cases.

%
%
\citeauthor{10.1145/3649506}~\cite{10.1145/3649506} provides a complete guide explaining how to use LLMs from various perspectives, including model selection, data consideration, and task specificity. 
It thoroughly investigates the details of pre-training and the significance of the training and test data and offers insights into tasks that require a rich knowledge base, natural language comprehension, generation capabilities, and other emergent features.
In particular, LLMs such as the Generative Pre-trained Transformer 3 (GPT-3), have gained attention due to their remarkable ability to acquire a broad spectrum of knowledge during pre-training and apply it to downstream NLP tasks \cite{ding2022gpt}, with significant potential for chatbots \cite{touvron2023llama,chiangvicuna,Dolly,taori2023alpaca,alpaca-lora,noauthor_falcon_nodate,noauthor_openai_nodate}.

\subsection{Leveraging LLMs for cybersecurity NLP}

Employing the capabilities of LLMs in ML tasks, specifically within the scope of cybersecurity, offers several possibilities \cite{zhang2023mlcopilot}.
Two problems that stand out are text classification, which categorizes text according to its relevance to a cybersecurity context, and NER, which recognizes specific cybersecurity entities in the text, for example, if it describes a vulnerability, which products are affected.
LLMs have shown outstanding effectiveness in many NLP tasks, including binary classification and NER \cite{min2023recent,KOCON2023101861}.

\subsection{OSINT for cybersecurity}
\label{RWo4cs}

In the CTI field, NLP tasks such as text classification and NER, which involve exploring various cybersecurity threats and concepts, are crucial.
In a previous study, SYNAPSE \cite{alves2021processing}, an OSINT processing pipeline was implemented to efficiently identify and concisely present various cybersecurity incidents to security analysts.
SYNAPSE was motivated by the results of previous works (e.g., \cite{alves2020follow,sabottke2015vulnerability}) that analyzed the completeness and timeliness of cybersecurity-related OSINT on Twitter.

SYNAPSE employed Support Vector Machines (SVM) \cite{cortes1995support} for binary classification, and a novel stream clustering approach to aggregate related tweets.
To improve SYNAPSE, \citeauthor{dionisio2019cyberthreat}~\cite{dionisio2019cyberthreat} designed a new tool that employs a CNN to detect cybersecurity-related texts gathered from Twitter and a BiLSTM network for performing NER on the detected tweets to identify the type of threat, affected products and other information.
That work was further extended using a multitask Deep Learning (DL) approach \cite{dionisio2020towards} that could simultaneously perform classification and NER.

Although there are many ML techniques for binary classification and NER~\cite{li2022survey,jehangir2023survey}, to the best of our knowledge, the architecture of \citeauthor{dionisio2019cyberthreat}~\cite{dionisio2019cyberthreat,dionisio2020towards} is recognized~\cite{altalhi2021survey} as the state-of-the-art for OSINT-based CTI extraction on Twitter data, reaching almost 95\% for different quality metrics.
This work also shows superior performance when compared with classifiers based on SVMs and Multi-Layer Perceptrons (MLPs), and several alternative NER approaches.
Due to this, we consider \citeauthor{dionisio2020towards}~\cite{dionisio2020towards} as the reference specialized model in our comparative analysis.

\subsection{Chatbots for Cybersecurity}
\label{RWcb4cs}

Although traditional and DL techniques have paved the way for advancements in cybersecurity text classification and NER, a new wave of NLP tools offers even more refined capabilities.
Among these tools, LLMs stand out for their exceptional capabilities in interpreting and generating human-like text.

The specific case of ChatGPT raised discussions and reviews on its potential applications in the cybersecurity field \cite{OKEY2023103476,gupta2023chatgpt,al2023chatgpt} and specific attack mitigation approaches have already been described \cite{MCINTOSH2023103424}.
Other works designed chatbots specifically to help cybersecurity analysis.
SecBot~\cite{franco2020secbot} is a cybersecurity-driven conversational chatbot that extracts information from a conversation to support cybersecurity planning and management.
It was developed and evaluated on a small dataset utilizing the Rasa framework \cite{conversational}, achieving 100\% accuracy in extracting the intent of attacks and associated named entities.

\subsection{Evaluation of LLM Chatbots}

Given the widespread attention LLM chatbots are attracting, several works tried to evaluate the strengths and limitations of the technology.
An analysis of ChatGPT's performance~\cite{KOCON2023101861,sun2023pushing} showed that despite its achievements, it frequently fell behind supervised baselines in various NLP tasks.
Several factors influence this, including limitations on the number of tokens, a misalignment with specific NLP tasks due to its generative nature, and challenges inherent to LLMs, such as hallucination, which involves making false positive predictions.
In their effort to optimize ChatGPT, the authors have proposed solutions, including multiple prompts, task-specific fine-tuning, and strategies to counter hallucination.
These methods were thoroughly tested across 21 datasets, covering ten critical NLP tasks, including NER.
The testing of the solutions resulted in significant performance improvements for ChatGPT, with instances where it outperformed state-of-the-art models in existing benchmarks \cite{sun2023pushing}.

The study by \citeauthor{megahed2023generative}~\cite{megahed2023generative} shows that ChatGPT excels in structured tasks like code translation and explaining established concepts.
However, it faces challenges when handling nuanced tasks such as recognizing unfamiliar entities and generating code from scratch.

Recent research findings show that the ChatGPT may encounter challenges and limitations in accurately identifying entities, including locations, names, and organizations.
\citeauthor{qin2023chatgpt}~\cite{qin2023chatgpt} present the results of their experiments on the performance of GPT-3.5, ChatGPT, and fine-tuned models on the multi-domain CONLL dataset for recognizing entities.
According to the findings reported in this paper, ChatGPT and GPT-3.5 achieved $\text{F}_1$ score of 53.7\% and 53.5\%, respectively.
\citeauthor{sun2023pushing}~ \cite{sun2023pushing} investigated the factors contributing to the sub-optimal performance of GPT-based chatbots in NLP tasks, such as NER.
They have identified several underlying causes and have proposed a set of generalized modules to mitigate these challenges in different NLP tasks.
\citeauthor{KOCON2023101861} \cite{KOCON2023101861} examined the capabilities of ChatGPT on a diverse set of subjective analytical tasks and objective reasoning tasks, revealing that when compared to state-of-the-art models, the average quality loss is about 25\% in zero-shot and few-shot settings.

As LLM-based chatbots have become widespread, concerns about their own vulnerabilities and their role as tools for cyber- attacks have increased~\cite{qammar2023chatbots}.
%
%
The usefulness of ChatGPT extends the time-to-conquer or delay attacker timelines, making it valuable for organizations seeking to enhance their cybersecurity posture \cite{mckee2023chatbots}.
Additionally, the performances of the ChatGPT and GPT-3 models are evaluated for vulnerability detection in code \cite{cheshkov2023evaluation}.
Based on a real-world dataset, this evaluation focuses on binary and multi-label classification tasks related to \emph{common weakness enumeration} of vulnerabilities.
The findings indicate that ChatGPT does not outperform the baseline classifier in classification tasks for code vulnerability detection \cite{cheshkov2023evaluation}.
More precisely, the performance of both GPT models was assessed by accuracy, precision, recall, $\text{F}_1$ score, and \emph{area under the curve}. 
The highest $\text{F}_1$ score of 0.67 was achieved using the text-davinci-003 model for binary classification. 
In contrast, the $\text{F}_1$ score of all ChatGPT models remained below 0.53 for multilabel classification.

Sentiment analysis plays a crucial role in extracting user opinions and emotions from textual data to assess threats.
Recent research has been directed toward developing sustainable strategies to diminish threats, vulnerabilities, and data manipulation within chatbots, ultimately improving the scope of cybersecurity. 
To achieve this objective, researchers created an interactive chatbot using the Bot Libre platform\footnote{\url{https://www.botlibre.com/}.} and placed it on social media platforms such as Twitter for the specific purpose of cybersecurity~\cite{arora2023developing}.
This study employs a sentiment analysis strategy by deploying chatbots on Twitter and subsequently analyzing Twitter data to anticipate forthcoming threats and cyberattacks.

\subsection{The Research Gap}

The reviewed papers in this section primarily focused on evaluating and testing the commercial ChatGPT model, with limited attention given to assessing open-source chatbot models in the context of cybersecurity applications.
Moreover, to the best of our knowledge, there is a notable absence of comprehensive comparative studies explicitly dedicated to open-source chatbots within the specialized field of OSINT-based CTI.
This gap in existing research underscores the need to examine open-source chatbot models, their effectiveness, and their potential contributions to enhancing cyber threat awareness and detection.
Although there are specialized state-of-the-art models, including DL models that excel at CTI binary classification and NER tasks~\cite{dionisio2019cyberthreat,dionisio2020towards,alves2021processing}, no comparative study has been conducted to determine whether LLM-based chatbots can compete with their performance.
If they can provide competitive results, CTI tools could be changed to integrate LLM chatbots into the OSINT processing pipeline, decreasing the tools' complexity and maintenance costs.
By employing general-purpose LLM chatbots, tasks related to CTI data collection, curation, labelling, and model training and updating would no longer be as necessary as they are for specialized models.

Our study aims to fill this gap by carefully comparing open-source and publicly available paid chatbots in the context of an OSINT-based CTI application, considering two downstream NLP tasks: binary classification and NER.

\section{LLM-based chatbots}
\label{LLM-based Chatbots}

LLM-based chatbots simulate human-like conversations with users through text or speech interaction.
They offer users more intelligent and contextually relevant responses to queries by utilizing LLMs' language comprehension and generation capabilities.
This section reviews the eight state-of-the-art commercial and open-source chatbot models available at the time of writing, namely, LLaMA, GPT4all, Dolly 2.0, Stanford Alpaca, Alpaca-LoRA, Vicunna, Falcon, and ChatGPT.
We used these models in our experimental evaluation (Section~\ref{Experimental Results}) to assess their effectiveness in detecting cybersecurity-related texts and identifying relevant entities.

\subsection{LLaMA}
LLaMA \cite{touvron2023llama} is a compilation of 7 to 65 Billion (B) parameter-based foundation language models.
These were trained on trillions of tokens, demonstrating the possibility of training cutting-edge models using only publicly accessible datasets. 
Their training method is comparable to that described in prior research \cite{brown2020language} and is influenced by the Chinchilla scaling laws \cite{hoffmann2022training}. 
Specifically, LLaMA-13B demonstrated superior performance compared to GPT-3 (175B) across a wide range of benchmarks, whereas LLaMA-65B exhibited similar performance levels to leading models, such as Chinchilla-70B and PaLM-540B. 
LLaMA models undergo training on substantial textual datasets by employing a conventional optimizer and large-scale transformers \cite{touvron2023llama}.

\subsection{Vicunna}
An open-source chatbot, Vicuna-13B \cite{chiangvicuna}, was developed by fine-tuning LLaMA with user-shared conversations gathered from the 70K ShareGPT.
In Vicuna, gradient checkpointing \cite{chen2016training} and flash attention \cite{dao2022flashattention} alleviate the memory demand.
The Vicuna report states that like other large language models, it has limitations. 
For example, it is not adept at tasks requiring logic or mathematics, and may have limitations in correctly identifying itself or assuring the factual performance of its outputs \cite{chiangvicuna}.


\subsection{GPT4all}
GPT4all \cite{anand2023gpt4all} utilizes LLaMA, which operates under a non-commercial license. 
The data for the assistant come from OpenAI's GPT-3.5-turbo, which has restrictions that prevent the development of models that directly compete with OpenAI in commercial applications.
GPT4all underwent several iterations with different versions featuring different parameter sizes. 
While preparing this article, the initial version had 7B parameters; however, the latest iteration used 13B.

\subsection{Dolly}
Dolly \cite{Dolly} operated by slightly modifying an open-source model with 6B parameters sourced from EleutherAI \cite{eleutherai}.
These modifications enabled Dolly to possess instruction-following capabilities, such as brainstorming and text generation, which were not initially present in the base model.
These modifications are implemented using the data from Alpaca \cite{taori2023alpaca}.
Subsequently, Dolly-v2-7b emerged as a highly advanced 6.9B parameter causal language model derived from EleutherAI's Pythia-6.9b.
Although Dolly-v2-7b may not be considered a state-of-the-art model, it exhibits instruc\-tion-following capabilities of remarkably high quality, which are not typically associated with its foundational model.

The most recent version available is Dolly-v2-12b \cite{noauthor_databricksdolly-v2-12b_2023}, a model with 12B parameters developed based on EleutherAI's Pythia-12b.
It has been finely tuned using a dataset called databricks-dolly-15k, which consists of an instruction corpus created by employees of Databricks.\footnote{https://www.databricks.com/blog/2023/04/12/dolly-first-open-commercially-viable-instruction-tuned-llm}
\subsection{Stanford Alpaca}
Stanford Alpaca \cite{touvron2023llama} is an instruction-following language model that is fine-tuned from Meta's LLaMA 7B model. 
It was instructed using 52k self-instruct-style demonstrations \cite{wang2022self}.
Alpaca displays several shortcomings in language models, such as hallucination, toxicity, and stereotypes. 
Specifically, hallucination appear to be a recurring issue in Alpaca \cite{taori2023alpaca}.
\subsection{Alpaca-LoRA}
This model \cite{alpaca-lora} reproduces the Stanford Alpaca results using low-rank adaptation (LoRA) \cite{hu2021lora}.
LoRA fine-tuning is a strategy for reducing memory requirements by employing a limited set of trainable parameters, known as adapters, rather than updating all model parameters, which remain constant.
\subsection{ChatGPT}
OpenAI's ChatGPT \cite{noauthor_openai_nodate} has generated substantial interest and sparked extensive discussion within the NLP community as well as in various other domains.
The lack of clarity regarding the training process and architectural specifics of ChatGPT poses a significant obstacle to both research endeavors and the advancement of open-source innovation within this domain.
Moreover, the distinction between the ChatGPT API and the web version model should be acknowledged.
Recent research shows considerable variability in the performance and behavior of GPT-3.5 and GPT-4 models over time \cite{chen2023chatgpt}. 
Another feature of generative AI models such as ChatGPT is that they do not produce repetitive responses to specific prompts \cite{megahed2023generative}.
\subsection{Falcon}
The Falcon family \cite{noauthor_falcon_nodate} consists of two primary models: Falcon-40B and its smaller counterpart, Falcon-7B. 
The Falcon-7B and Falcon-40B models underwent training on a corpus of 1.5 trillion and 1 trillion tokens, respectively \cite{noauthor_falcon_nodate}.
A feature of Falcon models is their utilization of multiquery attention \cite{shazeer2019fast}. 
In the vanilla multihead attention scheme, each head is associated with a query, key, and value. 
However, in the multiquery approach, a single key and value are shared across all heads.
\section{Chatbots evaluation}
\label{Chatbot Evaluation}
We divided this section into three topics relevant to the evaluation: dataset, experimental methodology, and experimental results evaluation criteria.
First, we introduce the Twitter dataset, which is the foundational source for generating subsequent prompts. 
The methodology subsection outlines the techniques and strategies used to assess and compare chatbot performances. 
Finally, in the evaluation criteria subsection, we explore the metrics and standards by which the chatbots are assessed.
\subsection{Dataset}
\label{sec:data}
We leveraged a comprehensive Twitter dataset made available by \citeauthor{alves2020follow}~\cite{alves2020follow} in their work to retrieve the Indicators of Compromise (IoCs) from Twitter OSINT. 
This dataset contains a combined total of 31281 tweets collected during two distinct periods: one from November 21, 2016, to March 27, 2017, and the other from June 1, 2018, to September 1, 2018. 
After collection, the tweets were filtered using specific keywords and manually labelled as positive or negative, considering their relevance to the cybersecurity of an IT infrastructure, thus creating labelled datasets suitable for supervised learning.
Later, \citeauthor{dionisio2019cyberthreat} labeled the dataset \cite{dionisio2019cyberthreat} for NER and republished it.

The dataset consists of 31281 tweet records, including the timestamp of the tweet, specific keywords found in the tweet, the original tweet, a pre-processed tweet cleaned from some special characters, a binary label marking the tweet as relevant for cybersecurity or not, and a string identifying the named entities in the pre-processed tweet.
Table \ref{tab:tweets} in \ref{AppendixB} shows representative examples of dataset records.

In this work, we used the pre-processed tweets, the cybersecurity relevance, and the sequences of named entity tags, to create a customized dataset for the binary classification and NER experiments through which the chatbots will be evaluated.
\subsection{Experimental methodology}
\label{Methodology}
As discussed in previous work \cite{alves2021processing,dionisio2019cyberthreat,dionisio2020towards}, the extraction of actionable CTI from OSINT entails three foundational tasks: text classification, extraction of pertinent information from textual sources, and the synthesis of the information gathered into concise summaries.
Considering the role of summarization to aggregate correlated threat indicators prior to dissemination, this study directs its attention towards evaluating and comparing chatbot performance within the two first tasks. 
These NLP tasks, being more amenable to LLM methodologies \cite{min2023recent}, are deemed as primary focal points for research within the scope of this study.

From a practical point of view, we interacted with the chatbots using Python language scripts to iterate through the dataset entries. 
For each entry, different prompts were used to address the two tasks of interest: deciding on the relevance of a tweet for cybersecurity using a binary classification formulation and extracting relevant information from the tweets using NER.

\subsubsection{Binary classification} 
\label{BC}
We designed prompts to determine whether tweets are related to the cybersecurity field, aiming to generate a response including `yes' or `no'. 
We focused on constraining the chatbot to provide concise yes or no answers without additional explanations. 
This restriction simplifies the process of extracting binary labels (0 or 1) from the responses.
For instance, in a scenario where the ``pre-processed tweet" column within our dataset contained the text ``cyber infosec kenno media SQL injection", to compose the specific prompts we desired, we added the phrase \textit{``Is the sentence"} at the beginning of each pre-processed tweet.
Then, we appended the phrase \textit{``related to cybersecurity? Please respond with yes or no."} at the end.
Consequently, the final prompt becomes: 
\textit{Is the sentence `threatmeter dos microsoft internet explorer 9 mshtml cdisp node::insert sibling node use-after-free ms13-0' related to cybersecurity? Just answer yes or no.} 

This process was applied to all the tweets in the ``pre-processed tweet" column to create a question dataset.
We administered three distinct tests, each consisting of the 31281 questions asked to chatbots.
\begin{itemize}
    \item {Test 1- Normal Dataset: In this test, we assess the performance of chatbots using the Twitter dataset described before, keeping the order of the rows unchanged. 
The objective is to assess how well chatbots can accurately identify tweets related to cybersecurity. 
 }
    \item {Test 2- Shuffled dataset:
To examine the impact of question order on the resulting answers, the question dataset was shuffled and provided as input to the models. }
    \item {Test 3- Isolated Prompt Testing:
In this test, we indicate that each question was regarded as an isolated prompt without considering the context of past interactions.
This test aims to clear the chatbot's history and context between consecutive questions during testing.
It allows for clean and unbiased interaction with the chatbot for each question.
For this test, we employ a methodology that closes the chatbot session after each question and initializes a fresh session for the subsequent question.
This ensures the chatbot does not retain any knowledge or bias from previous interactions, thus providing a fair evaluation of its performance on individual questions.}
\end{itemize}

We conducted research on two models of commercial ChatGPT, namely GPT-3.5-turbo and GPT-4, using versions that were available on August 5th, 2023.
The API provided by OpenAI was used to send requests to ChatGPT for each tweet in the dataset.
Various parameters can be configured by utilizing the ChatGPT API.
One particularly influential parameter is referred to as ``temperature".
This parameter governs the level of creativity or randomness of the generated text.
A higher temperature setting (e.g., 0.7) produces a more varied and imaginative output, whereas a lower setting (e.g., 0.2) yields a more predictable and concentrated output.
We adjusted the temperature parameter to 0.2 to reduce output randomness while leaving the other parameters at their default values.

\subsubsection{Named entity recognition}
\label{NER}
The assessment of chatbot performance in NER involves two distinct methodologies.

The initial approach, which is called Entity-Specific Prompting (ESP), was chosen based on a careful evaluation of the various experimental results.
It was evident from these experiments that attempting to extract all required entities from a single tweet using a single question yielded significant misrecognition, rendering the approach impractical.
Consequently, we decided on a more precise strategy involving the creation of specific prompts for each entity within each tweet.
After sending several prompt requests for each entity to the chatbots, we found that the organization names and product version entities were the most extractable by chatbots.
The revised approach focuses on two specific entities: organization names (B-ORG) and product versions (B-VER).
We selected these entities for extraction by interacting with ChatGPT-3.5-turbo, ChatGPT-4, GPT4all, and Dolly models.
Since only 11074 out of 31281 questions in the dataset had been tagged with NER labels, we limited our analysis for the NER task to these 11074 tweets.
In our experiment, we employed various prompts to choose the proper one; ultimately, we selected the following prompts to identify the organization name and product version:

\begin{itemize}
\item{\textit{Find the name of organizations in the following sentence: `threatmeter dos microsoft internet explorer 9 mshtml cdisp node::insert sibling node use-after-free ms13-0'. 
Give the shortest answer, and only use sentence segments in your response.}}
\item{\textit{Find only product version numbers without any product, vulnerability, and company names in the following sentence: `threatmeter dos microsoft internet explorer 9 mshtml cdisp node::insert sibling node use-after-free ms13-0'. 
Give the shortest answer, and only use sentence segments in your response.}}
\end{itemize}

The second approach, which is called Guide-Line Prompting (GLP), exclusively employed for ChatGPT-4, involves a comprehensive specification of all entities within the `GUIDELINES\_PROMPT' section.
This approach attempts to extract seven entities in a single prompt and considers only the ChatGPT model because the GUIDELINES\_PROMPT feature is absent in open-source models.
In this guideline, we include two examples of tweets from 11074 NER-tagged tweets in the dataset, each annotated with their respective entities.
In addition, we have included the output format as a means to direct ChatGPT's response for subsequent processing.
We then sent a dedicated prompt for each of the 11074 tweets and systematically covered all the extracted entities.
The prompt guideline is given in \ref{AppendixA}.

\subsection{Evaluation criteria}
\label{Evaluation Criteria}
This section describes our evaluation criteria to assess chatbot models' performance in addressing binary classification and NER tasks. 
Our evaluation methodology focuses on performance and quality.

\subsubsection{Performance}
To assess the selected chatbots' performance, we used the $\text{F}_1$ score, which is a metric that computes the harmonic mean of Precision and Recall to measure the binary classification performance of a model.
It strikes a balance between the proportion of true positive results in all positive predictions and the proportion of true positive results in all actual positives. 
The $\text{F}_1$ score is given by
\begin{equation}
    \text{F}_1 = 2 \times \frac{\text{Precision} \times \text{Recall}}{\text{Precision} + \text{Recall}}
\end{equation}
where Precision is the ratio of correctly predicted positive observations to the total predicted positives, given by
\begin{equation*}
    \text{Precision}=\frac{TP}{TP + FP},
\end{equation*}
and Recall measures the ratio of correctly predicted positive observations to all observations in the positive class:
\begin{equation*}
    \text{Recall}=\frac{TP}{TP + FN}.
\end{equation*}


\subsubsection{Quality}
For the classification task, the response quality of the models relates to their ability to provide precise yes or no answers to the prompts they receive.
The assessment is based on three distinct response modes that the models eventually produced: no response, correct response, and implicit response.
\begin{itemize}
    \item{No response: Instances in which chatbot models failed to respond.}
    \item{Precise response: The questions to which the chatbot models provided accurate answers were precisely aligned with the desired yes or no response.}
    \item{Implicit response: Answers in which the chatbot's response did not explicitly mention yes or no. 
However, careful inference from the generated answers revealed that the intended response was either yes or no.}
\end{itemize}
By analyzing these distinct response modes from the output files, we gain valuable insights into the performance and capabilities of the evaluated chatbot models to effectively detect and address cybersecurity-related questions. 
The experimental results section provides a detailed explanation of the experimental findings, focusing on the quality and accuracy evaluation criteria.


\section{Optimal utilization strategies of chatbots}
\label{Strategies for Optimal Utilization}
The fundamental principle behind creating prompts is ensuring clear and precise instructions.
Prompt engineering~\cite{liu2023pre} is crucial for optimizing the performance of chatbot models by enhancing the clarity and specificity of the given instructions.
Thorough prompt design and testing improve the ability of the model to comprehend requests, making it a more effective tool for generating desired outcomes. 
These approaches enhance the possibility of directing the model to the desired output while decreasing the chances of receiving irrelevant or incorrect responses.
\subsection{Prompt enginering approaches}
We devoted considerable attention to formulating suitable prompts to maximize accurate and relevant responses from the chatbots during our experiments.
The objective was to design prompts that effectively allow models to capture the essence of the CTI task in the information present in tweets.

Following the best practices for prompt engineering [58], we progressively refined our prompts using two approaches.
For NER, we considered first a prompt guideline template approach but with poor results (prompt template is shown in \ref{AppendixA}).
We used a zero-shot methodology for binary classification, considering that LLMs are excellent zero-shot reasoners \cite{kojima2022large}.
Since this approach produced very good results, for the NER tests, we transformed the prompt guideline into a sequence of zero-shot questions, one for each entity we aim to extract.

To leverage the zero-shot capability of LLMs, we explored the assumption that chatbots perfectly represent the cybersecurity concept.
Following this assumption, we designed prompts by starting the question with information on the cybersecurity issue and finishing it with precise instructions on the expected outcome.

The process of selecting the final prompt was iterative and relied on a few trial and error cycles, using the principles described before.
Using a sample of dataset entries, we progressively queried the chatbots and evaluated the answers until consistent answers were achieved concerning the instructions given.
This refinement cycle resulted in the determination of the final prompt.

By formulating questions such as \textit{Is the sentence `vuln oracle java se cve-2016-5582 remote security vulnerability' related to cybersecurity? Just answer yes or no.}, we aimed to elicit yes or no responses from the chatbots.
Using spaces and adequately employing the apostrophe (') played a significant role in clarifying the prompt.

This methodology allowed us to assess the chatbot's capabilities to detect cybersecurity concepts and their ability to generate meaningful and contextually appropriate responses.
Having formulated the desired prompt, we automated the process of sending the entire set of questions to chatbots.

\subsection{Text length control}
Chatbots can produce text of varying lengths based on specific tasks.
In our experiments, minimizing the number of answer tokens was essential because of the high volume of questions and the time required for the model to answer each question.

The parameter that defines the length of the answer or the number of tokens in local chatbots is represented as \emph{N\_predict}.
It plays a critical role in significantly reducing the execution time for answering questions across all open-source chatbot models.
To optimize the execution time, we advise setting a parameter that controls the response length to the smallest suitable value based on a specific task.
After some initial experimentation, we consistently set the value of \emph{N\_predict} to 15 across all open-source models.

In the context of ChatGPT-3.5-turbo and ChatGPT-4, the \emph{max\_tokens} parameter constrains the length of the responses generated by the model.
This is achieved by establishing a predetermined upper limit for the number of tokens that can be words or characters within the generated output.
Using more extended responses in ChatGPT-4 can increase token consumption, potentially increasing the usage costs.
After some initial experimentation with prompts with extreme lengths, the value 70 was assigned to \emph{max\_tokens} in the experimental tests.
It is worth noting that the chosen token length of 15 for the open-source chatbots applies exclusively to the generated answers.
By contrast, the selected token length of 70 for ChatGPT models encompasses questions and answers.

The careful utilization of the \emph{N\_predict} and \emph{max\_tokens} parameters is of utmost importance, as a low setting may lead to truncation of the response, potentially producing incomplete or nonsensical answers.
Balancing the desired response length with the need for completeness and coherence is a crucial factor to consider.

\section{Results and Discussion}
\label{Experimental Results}
In this section, we present the results of the empirical assessment of chatbots’ thorough evaluations of their capabilities across multiple dimensions. 
First, we discuss the evaluation of binary classification, focusing on how chatbots classify user inputs proficiently. 
Next, we present the evaluation of NER tasks by examining the effectiveness of chatbots in identifying and classifying entities present within user inputs.  
The collective findings from these experiments offer comprehensive insights into chatbots’ operational strengths and potential areas for improvement.

It is essential to highlight that we ultimately elaborate on the common and default parameters shared by all open-source chatbot models. 
In most of these models, the maximum size of the context window\footnote{Upper limit for the range of tokens considered to provide an answer.} and its default value are set to 512.
An exception is the Dolly model, which has a maximum context window size of 1024.
However, the commercial ChatGPT model exhibits variability across its final versions, each with a distinct context window size. 
For instance, ChatGPT-4 is available in two versions with window sizes of 8k and 32k, whereas ChatGPT-3.5-turbo is available in 4k and 16k versions.
In our experiments, we utilized a server equipped with multiple GPU units, including an NVIDIA A30 GPU (memory capacity: 24,576 MiB) and an NVIDIA RTX A6000 GPU (memory capacity: 49,140 MiB), with 264 GB of RAM.

\subsection{Evaluation of binary classification}
We present the results obtained from chatbots, focusing on two critical dimensions: quality and accuracy.
\begin{table*}
\centering
\renewcommand{\arraystretch}{1.4}
\caption{Accuracy of chatbot models for cybersecurity binary classification}
\label{tab:accuracy}
\begin{tabular}{ccccccc} 
\hline
\rowcolor[rgb]{0.902,0.902,0.902} Model & Test Number & \begin{tabular}[c]{@{}>{\cellcolor[rgb]{0.902,0.902,0.902}}c@{}}  Parameters\end{tabular} &Precision &Recall&$\text{F}_1$ score                   & Execution Time  \\ 
\hline
ChatGPT-3.5-turbo (16k context) \cite{noauthor_openai_nodate}            & Test 1       & 175B                                                                                                     &0.9570 &0.9280     & 0.9431                       & 11h 23m         \\
ChatGPT-3.5-turbo (16k context) \cite{noauthor_openai_nodate}           & Test 2       & 175B                                                                                                     &\textbf{0.9700} &0.9200     & \textbf{0.9489}                       & 11h 23m         \\
ChatGPT-3.5-turbo (16k context) \cite{noauthor_openai_nodate}            & Test 3       & 175B                                                                                                     &- &-     & UECH                  & -         \\
ChatGPT-4 (8k context) \cite{noauthor_openai_nodate}                      & Test 1       & 1.7T                                                                                                      &0.9580 &0.9240    & 0.9410                       & 11h 50m         \\ 
ChatGPT-4 (8k context) \cite{noauthor_openai_nodate}                      & Test 2       & 1.7T                                                                                                      &0.9590 &0.9230    & 0.9403                       & 11h 43m        
\\
ChatGPT-4 (8k context) \cite{noauthor_openai_nodate}                     & Test 3       & 1.7T                                                                                                      &- &-    & UECH                      & -        \\
\hline
GPT4all \cite{anand2023gpt4all}                                & Test 1       & 13B                                                                                                      &0.9490 &0.8630  & 0.9049                     & 132h 05m        \\
GPT4all                                 & Test 2       & 13B                                                                                                     &0.9490 &0.8410  & 0.8927                     & 132h 02m        \\
GPT4all                                 & Test 3       & 13B                                                                                                      &0.9470 &0.8280  & 0.8844                     & 136h 05m        \\ 
\hline

Dolly 2.0 \cite{Dolly}                             & Test 1       & 7B                                                                                                 &0.8890 &0.8000        & 0.8470                     & 10h 38m         \\
Dolly~2.0                               & Test 1       & 12B                                                                                                       &0.9470 &0.7900 & 0.86120                     & 10h 16m         \\
Dolly~2.0                               & Test 2       & 12B                                                                                                     &0.9480 &0.7910   & 0.8631                     & 10h 00m~      
\\
Dolly~2.0                               & Test 3       & 12B                                                                                                        & -                &- &-     & LET ~      
\\ 

\hline

Falcon \cite{noauthor_falcon_nodate}                                 & Test 1       & 7B                                                                                                   &0.8120 &0.8500      & 0.8304                       & 16h 02m         \\
Falcon                                  & Test 1       & 40B                                                                                                      &0.8980 &0.8200  & 0.8511                       & 54h 03m     
\\
Falcon                                  & Test 2       & 40B                                                                                                      &0.8990 &0.8080  & 0.8502                       & 54h 55m     
\\
Falcon                                  & Test 3       & 40B                                                                                                      &0.8990 &0.7880  & 0.8330                       & 71h 10m     
\\ 

\hline
Alpaca-LoRA \cite{alpaca-lora}                                 & Test 1       & 65B                                                                                                    &0.8980 &0.7940    & 0.8477                     & 10h 12m         \\
Alpaca-LoRA                                   & Test 2       & 65B                                                                                                  &0.8990 &0.8000      & 0.8451                     & 10h 44m         \\

Alpaca-LoRA                                   & Test 3       & 65B                                                                                                     &0.8980 &0.7610   & 0.8241           & 11h 20m         \\

\hline

Stanford Alpaca \cite{taori2023alpaca}                        & Test 1       & 7B                                                                                                    &0.2260 &0.5000     & 0.3112                     & 13h 03m         \\
Stanford Alpaca                         & Test 1       & 13B                                                                                                    &0.3240 &0.6000    & 0.4209                     & 13h 21m         \\
Stanford Alpaca                         & Test 1       & 30B                                                                                                     &0.6980 &0.6050   & 0.6415                     & 15h 48m         \\ 
Stanford Alpaca                         & Test 2       & 30B                                                                                                    &0.6990 &0.5920    & 0.6401                     & 15h 04m         \\ 
Stanford Alpaca                         & Test 3       & 30B                                                                                                      &0.6990 &0.5810  & 0.6395                     & 16h 18m         \\ 

\hline

Vicuna \cite{chiangvicuna}                                  & Test 1       & 13B                                                                                                    & 0.4390&0.3100    & 0.3611 & 11h 23m         \\ 

\hline

 \citeauthor{dionisio2020towards}~ \cite{dionisio2020towards}                                     & Test 1       & -                                                                                                         &0.9570 &\textbf{0.9363} & 0.9470                      & 00h 43m             \\

\hline

\multicolumn{7}{l}{* LET: Long Execution Time   ~~~~~~~~~~~~~~~~~ * UECH : Uncertainty of Erasing Conversation History}
\end{tabular}
\end{table*}
\subsubsection{Performance}
\begin{figure}[!t]
    \centering
    \includegraphics[width=\columnwidth]{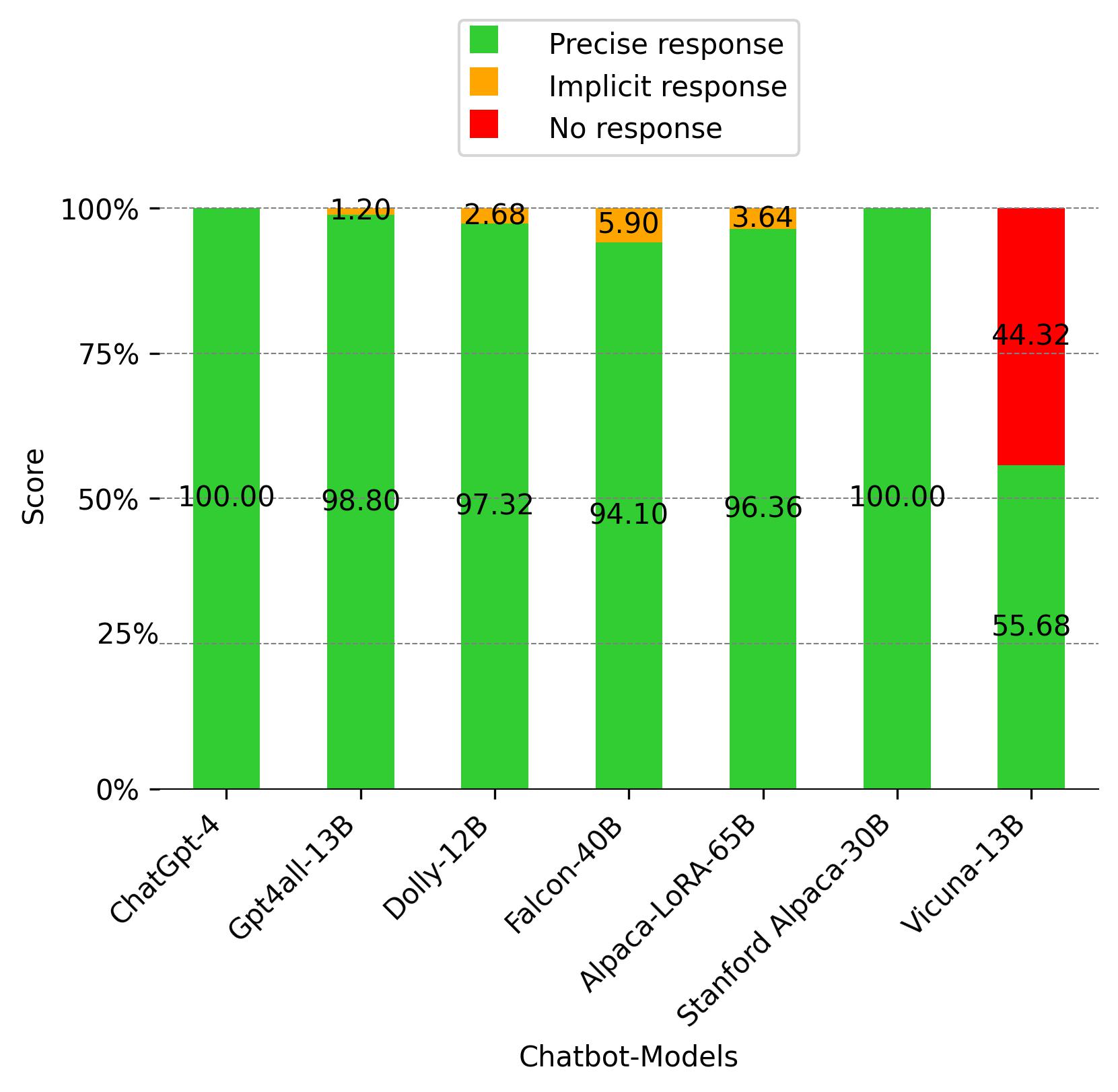}
    \caption{Comparison of Chatbot model response modes -Test 1}
    \label{fig: response}
\end{figure}

In terms of quality, different categories are shown in different colors.
The presence of red in \autoref{fig: response} highlights a noteworthy observation regarding the unanswered questions.
Precise responses are visually represented by the green bars on the chart, effectively indicating the successful accomplishment of the intended goal.
The orange bars in the diagram represent acceptable answers with minor imperfections, showing chatbots' capacity to offer responses that were contextually in line with the desired result, albeit with slight deviations.
The Vicuna model encountered significant challenges in answering 17420 questions despite our efforts to use iteration loops to generate answers.
To address these unanswered questions, we included them in the loop until the model could respond. 
However, a considerable number of questions have become trapped in an infinite loop and remain unanswered.
This problem is unique to the Vicuna chatbot model, with no other models showing it.
As shown in \autoref{fig: response}, ChatGPT and Stanford Alpaca exhibited excellent quality, with all 31281 questions consistently beginning with yes or no.
By contrast, the GPT4all model achieved a slightly lower rate, with 98.80\% of its responses starting with either yes or no.
For the Dolly and Falcon models, the rates are 97.32\% and 94.10\%, respectively.
The GPT4all model delivered conceptual yes or no responses to 378 questions, accounting for 1.2\%, whereas Dolly produced responses for 839 questions, constituting 2.68\%.
Interestingly, however, neither of these models explicitly used the words yes or no in these conceptual answers.
For example, the conceptual answers were: \textit{it is related to cybersecurity}, or \textit{the sentence is not related to cybersecurity.}

In terms of accuracy, we explain the results of the tests conducted on chatbot models. 
\autoref{tab:accuracy} provides an overview of the tests conducted for each chatbot and its version. 
Moreover, we recorded the execution time of the models' responses for all the questions.
It includes details regarding the number of parameters on which the model was trained, achieved $\text{F}_1$ score, precision, recall, and execution time for each model.
The confusion matrices corresponding to each model are provided in \ref{AppendixC}.
By analyzing the $\text{F}_1$ score values in Test 1, we can assess the effectiveness of the models in accurately responding to the given questions.
Based on the results presented in \autoref{tab:accuracy}, it is evident that the GPT4all achieved the highest accuracy among the open-source models, as indicated by its $\text{F}_1$ score of 0.90. 
The Dolly model has an accuracy of 0.86, Falcon of 0.85, Alpaca-LoRA of 0.84, and the Stanford Alpaca model has a score of 0.64.
Although GPT4all achieved higher accuracy among open-source chatbots, it is noteworthy that the commercial ChatGPT models (GPT-4 and GPT-3.5-turbo) achieved an $\text{F}_1$ score of 0.94.
ChatGPT-3.5-turbo with a 16k window context size achieves the same $\text{F}_1$ score as ChatGPT-4 with an 8k context window size.
These results highlight the better accuracy of the GPT4all and ChatGPT models, emphasizing their effectiveness for this particular task.

Based on the shuffled dataset test, the GPT4all model achieved an $\text{F}_1$ score of 0.89\%, whereas the Dolly model attained an $\text{F}_1$ score of 0.86\%.
 
This indicates a slight decrease in accuracy of approximately 1\% for the GPT4all model compared with the first test, which can be considered insignificant.
It is essential to mention that shuffling the prompt does not affect the accuracy of the ChatGPT, Falcon, Stanford Alpaca, and Alpaca-LoRA models, as the $\text{F}_1$ score is equal to that of the first test.

Upon applying the isolated prompt test to the dataset, the $\text{F}_1$ score for the GPT4all model attained a value of 0.88. 
This test results in a 2\% decrease compared with that of Test 1.
Conducting the test for the Stanford Alpaca model did not result in a significant reduction in accuracy.
When testing the Alpaca-LoRA model, we observed a two percent reduction in the $\text{F}_1$ score of 0.82.
The result of this test on the model Falcon 40B is two percent less than the first test, and it is equal to 0.83.
We ignored this test on Dolly because it was time-consuming and beyond the available time and resources.
For ChatGPT, we used the available API to send requests. 
In this case, Test 3 is not feasible because isolated prompt functionality is not available, i.e., the conversation history of the model cannot be completely reset, and we have no means to reinitialize the model for each prompt.
It is worth noting that running this test increases the execution time because of the need to launch the model for each question.
Nevertheless, the results obtained by the other models indicate that we should not expect a significant decrease in accuracy.

Chatbots are based on a varying number of parameters.
This study involved models ranging from 7B to 65B in parameter counts. 
Based on our experiment, the number of parameters significantly influences the effectiveness of the models in answering questions. 
Generally, models possessing 7B parameters exhibited a comparatively lower performance \cite{bi2024deepseek} and, based on our experiments, failed to provide binary ('yes' or 'no') answers.
Remarkably, GPT4all, with 7B parameters, faced limitations in providing yes or no responses to every question.
In response to each question, the output consisted solely of explanations, leading to a significant expenditure on human effort to discern whether the response was yes or no, thus preventing the automated processing of the answers. 
Consequently, we were unable to calculate the $\text{F}_1$ score for the 7B parameter model.
In addition, two other models, Dolly and Stanford Alpaca, both of which have precisely 7B parameters, have lower $\text{F}_1$ score than the models with 13B parameters, as shown in \autoref{tab:accuracy}.



\subsubsection{Execution time}
In our findings, the execution times of the models varied significantly.
The implementation process involves running LLMs on a GPU server, which requires several days of continuous execution.
GPT4all took the longest, with a total execution time of 132 hours and five minutes. 
Falcon followed with 54 hours and 3 minutes. Stanford Alpaca completed its tasks in 15 hours and 48 minutes, while Dolly had a slightly shorter execution time of 13 hours and 38 minutes. 
ChatGPT-4 was more efficient, ranking third with an execution time of 11 hours and 50 minutes. 
The fastest among them was Alpaca-LoRA, with an execution time of 10 hours and 12 minutes.
The models with versions of 30B and 65B parameters exhibit longer execution times when answering questions. 
An impressive aspect of the Alpaca-LoRA model with 65B parameters is that despite its significantly larger parameter count than Dolly, which has 7B and 12B parameters, both models exhibit equal execution times and $\text{F}_1$ score.

The last row of \autoref{tab:accuracy} is devoted to the multitask model \cite{dionisio2020towards} discussed in Section~\ref{RWo4cs} of the related work.
Since this model achieves the best performance in OSINT-based CTI extraction, it was selected as the reference specialized model to evaluate the chatbots.
Furthermore, the Twitter dataset employed for both this model and the chatbots is identical, guaranteeing a fair evaluation under comparable conditions.

This model is a Bidirectional Long Short-Term Memory (BiLSTM) trained for binary classification tasks and NER.
It achieved an accuracy of 0.94, equivalent to the ChatGPT model's accuracy.
However, the execution time is only 43 minutes, which is much shorter than the 11+ hours required by ChatGPT.
Based on the experimental results, it is evident that ChatGPT, with a context window size of 512, outperforms the open-source models in terms of the $\text{F}_1$ score. 
One key contributing factor to this accuracy gap is the significantly larger context window size employed by the ChatGPT.

\subsection{Evaluation of NER}
\autoref{tab: GPT-4 NER accuracy} shows the results for identifying the `product version' entity, which is 0.43 of the $\text{F}_1$ score, based on 11074 requested questions. 
This $\text{F}_1$ score exceeds the corresponding value for identifying the `organization name' entity.
Our analysis for the ESP approach is based on the ChatGPT version released on July 13th, 2023.
The results in the last line of \autoref{tab: GPT-4 NER accuracy} indicate that the calculated $\text{F}_1$ score for 11074 questions is unexpectedly low, reaching 0.10 accuracy.
The findings demonstrate a significant deviation from the current NER outcomes as presented in the \cite{dionisio2020towards} study, which achieved an $\text{F}_1$ score of 0.94.
Our GLP approach evaluation is based on the ChatGPT released on August 2, 2023.

A representative example of the responses generated by each model is provided in \autoref{tab: prompt reponse}.
In the two NER prompts mentioned in Section \ref{Methodology}, the annotations in the dataset show `Microsoft' as an organization entity and `9' as a product version.
The phrase \textit {`without any product, vulnerability, and company names'} in the version prompt plays a crucial role in constraining the models' interpretation of version numbers. 
ChatGPT-4 demonstrated precision in identifying `Microsoft' as B-ORG and `9' as B-VER, accurately matching the annotations in the dataset. 
By contrast, the other models exhibited varying degrees of accuracy, failing to reach the level of correctness achieved by ChatGPT-4 in these specific instances.
\definecolor{Mercury}{rgb}{0.901,0.901,0.901}
\definecolor{Mercury}{rgb}{0.901,0.901,0.901}
\begin{table*}
\centering
\renewcommand{\arraystretch}{1.7}
\caption{Comparison of NER task accuracy achieved by the ChatGPT-4 (8k context) model using two different approaches.}
\label{tab: GPT-4 NER accuracy}
\begin{tblr}{
  colspec={lccc}, 
  row{even} = {c},
  row{1} = {Mercury},
  row{3} = {c},
  cell{1}{1} = {c},
  cell{1}{2} = {c},
  cell{1}{4} = {c},
  cell{1}{5} = {c},
  cell{1}{6} = {c},
  hlines,
}
 Approach    & Number of Questions &  Entity          & $\text{F}_1$ score & Execution Time \\
ESP & 11074                       & Organization    & 0.36      & 4h 02m        \\
ESP & 11074                      &     Version & 0.43      & 4h 23m        \\
GLP & 11074                      & All entities & 0.10     & 3h 09m        
\end{tblr}
\end{table*}

\begin{table*}
\centering
\renewcommand{\arraystretch}{1.4}
\caption{Illustrative ChatGPT responses demonstrate NER identification for two distinct prompts.
}
\label{tab: prompt reponse}
\begin{tabular}{ccc} 
\hline
\rowcolor[rgb]{0.902,0.902,0.902} Model & Entity       & Models~\textcolor[rgb]{0.216,0.255,0.318}{Responses~}            \\ 
\hline
ChatGPT-4 (8k context)                                  & Organization & Microsoft                                                        \\
ChatGPT-4  (8k context)                                   & Version      & 9                                                                \\ 
\hline
GPT-3.5-turbo                           & Organization & The name of organizations in the given sentence is "Microsoft``.  \\
GPT-3.5-turbo                           & Version      & 9, ms13-0                                                        \\ 
\hline
Dolly 2.0                               & Organization & Microsoft Internet Explorer 9                                    \\
Dolly 2.0                               & Version      & ms13-0 is 9.0.8112.16421                                         \\ 
\hline
GPT4all                                 & Organization & Microsoft, Mozilla (Firefox), and Google Chrome                  \\
GPT4all                                 & Version      & 546                                                             \\
\hline
\end{tabular}
\end{table*}
\subsection{Challenges and limitations}
\label{challenges}
Chatbots present opportunities in numerous applications but also face challenges and limitations that hinder their effective utilization.
In this study, we outline several challenges and limitations encountered during our experiments, which are detailed in this section.
Additionally, we faced specific limitations when applying chatbots to the specialized CTI tasks, leading to issues with timeliness that are elaborated upon in Section \ref{BC:challenges}.
A common challenge to binary classification and NER was the generation of effective prompts, which required progressive refinement until satisfactory performance was achieved.
This section explores the challenges and limitations encountered in the binary classification and NER tasks.
\subsubsection{Binary classification}
\label{BC:challenges}
In Section \ref{BC}, we discuss how each chatbot model displayed unique response behaviors when answering questions.
This necessitated a cleaning step after collecting responses to ensure answer consistency, particularly because our goal was to obtain a precise binary (yes or no) response. 
However, some responses implied a `no' or a `yes' without explicitly using these words. 
For instance, a model might answer the question, `This is not related to cybersecurity'.
Consequently, it is crucial to review and validate the answers in the output file. 
Our evaluation process involved two key steps to ensure reliability and accuracy. 
First, we implemented an automated validation method for each output file to confirm the presence of `yes' or `no' responses at the beginning of the responses. 
This helped to filter out potentially incorrect answers, such as those lacking an explicit `no' or `yes'. 
Second, we conducted a thorough manual review of each response to confirm its accuracy and alignment with the expected answers. 
In cases such as the aforementioned example, we manually annotated the response with `no' to correctly classify it. 
This comprehensive manual validation adds a crucial layer of scrutiny, bolstering the accuracy and reliability of the results. 
While this comprehensive validation process enhances the accuracy of our binary classification, it also introduces another challenge: the need for timely response processing.
Timeliness is a critical factor in the use of chatbot models for binary classification tasks, particularly in CTI applications. 
The demand for real-time processing is essential, as any delay in classifying responses can significantly impact decision-making. 

\subsubsection{Named entity recognition}
When employing chatbots in NER tasks for cybersecurity purposes, we observed various limitations, mainly in providing precise and relevant results. 
While chatbots powered by pre-trained language models excel at understanding natural language and utilizing general knowledge, they frequently encounter challenges when dealing with domain-specific precise entity recognition \cite{10.1145/3649506}.
This shortcoming is attributed mainly to the intrinsic complexities of NER, which demand a profound understanding of context, domain knowledge, and syntactic intricacies.

A recurrent issue in chatbot-assisted NER is the generation of unspecific answers that fail to accurately identify precise entities in a given sentence. 
This often leads to generalized responses that lack the precision essential for obtaining reliable NER results. 
Consider this prompt as an example: \textit{Find the name of organizations in the following sentence:
`senator calls on us government to start killing adobe flash now tripwire’. 
Give the shortest answer, and only use sentence segments in your response.}
The ChatGPT-4 response was `US Government, Adobe, Tripwire'.
Such challenges can be traced back to factors such as the limitations inherent in the underlying language models or the absence of dedicated fine-tuning tailored to NER tasks.
Another phenomenon, hallucination, as discussed in Section \ref{RWcb4cs}, also arises from these compounded challenges.
As an example of precise hallucination, 
GPT4all mistakenly extracted `546' as a product version, the number that was not seen at all in the product version prompt, which is mentioned in Section \ref{NER}.

\subsection{Discussion}
\label{Discussion}
The evaluation of chatbot models involves two critical aspects.
First, it involves a deep understanding of effective methods for interacting with the models.
This includes considering timeliness, which is particularly important when integrating chatbots into real-time systems such as those connected to Twitter like SYNAPSE \cite{alves2021processing}. 
Second, evaluation requires the skill of writing structured and clear prompts for chatbots, ensuring that they produce precise and relevant responses.
The ability to compose clear and concise prompts that guide the chatbot to provide an anticipated response is crucial.
The preceding sections explain that exercising control over the response length is crucial. 
Long answers not only extend execution times but also impose an additional workload on human resources for response validation. 
On the other hand, those responses might not contain enough meaningful content.

Automated interaction with chatbots for tasks such as binary classification and NER is another scope to enter that requires specific requirements.
This includes the ability to automatically generate prompts that are contextually relevant and precise. 
The system must also interpret and adapt to varying response formats and manage the complexities of the different types of data inputs. 
Additionally, the chatbot's ability to handle ambiguity in natural language while maintaining the accuracy of its responses is essential. 
This requires sophisticated algorithms that are capable of understanding subtle differences in language and context.

Although our evaluation primarily focused on a specific Twitter dataset, it is worth noting that numerous other OSINT resources, such as blog posts and security forums (even in the dark web), remain unexplored. 
Moreover, it is important to note that a manual review of chatbot responses that follow the automated checking phase may introduce potential human errors into the assessment process.
Further, such verification is time-consuming and thus not feasible in the day-to-day operation of a security operating centre.


\section{Conclusion}
\label{Conclusion}
We assess the capabilities of open-source and paid LLM-based chatbots to recognize cybersecurity-related tweets and extract pertinent information from these.
Both types of models can perform similarly to specialized models trained specifically for the binary classification task of identifying cybersecurity-related tweets, often achieving the same level of performance.
On the contrary, the chatbot models' performance is still very poor on named entity recognition to extract security elements from tweets.
Even when training on vast datasets, these models did not perform comparably to specialized models on the test data.

Our results highlight the need for further research and refinement in the application of LLM chatbot models to extract threat indicators from open-source intelligence.
Although open-source and paid chatbots compared evenly with specialized trained models on cybersecurity binary text classification, they fell below acceptable performance on named entity recognition. 
Moreover, they can not compete with specialized models on timeliness and cost.

Based on our study, we identify various possibilities for future work based on the following research questions:

\begin{enumerate}
    \item How can LLM chatbots be further optimized for cost-effective real-time CTI detection on social media platforms?
    \item How to improve the NER capability of chatbots for the extraction of indicators of compromise?
    \item How can cybersecurity specialists' feedback be used to increase the efficiency and cost-effectiveness of open-source chatbots?
\end{enumerate}


\section*{Acknowledgement}
This work is funded by the European Commission through the SATO Project (H2020/IA/957128) and by FCT through the LASIGE Research Unit (UIDB/00408/2020 and UIDP/00408/2020).
\balance
\bibliographystyle{elsarticle-num-names}
\bibliography{cas-dc-template}
\pagebreak
\begin{appendices}

\renewcommand{\thesection}{Appendix \Alph{section}}
\renewcommand{\thefigure}{A\arabic{figure}}
\setcounter{figure}{0}
\renewcommand{\thetable}{B\arabic{table}}
\setcounter{table}{0}
\definecolor{Alto}{rgb}{0.862,0.858,0.858}

\onecolumn

\section{}
\label{AppendixA}
\begin{figure*}[]
    \centering
    \includegraphics[width=\linewidth]{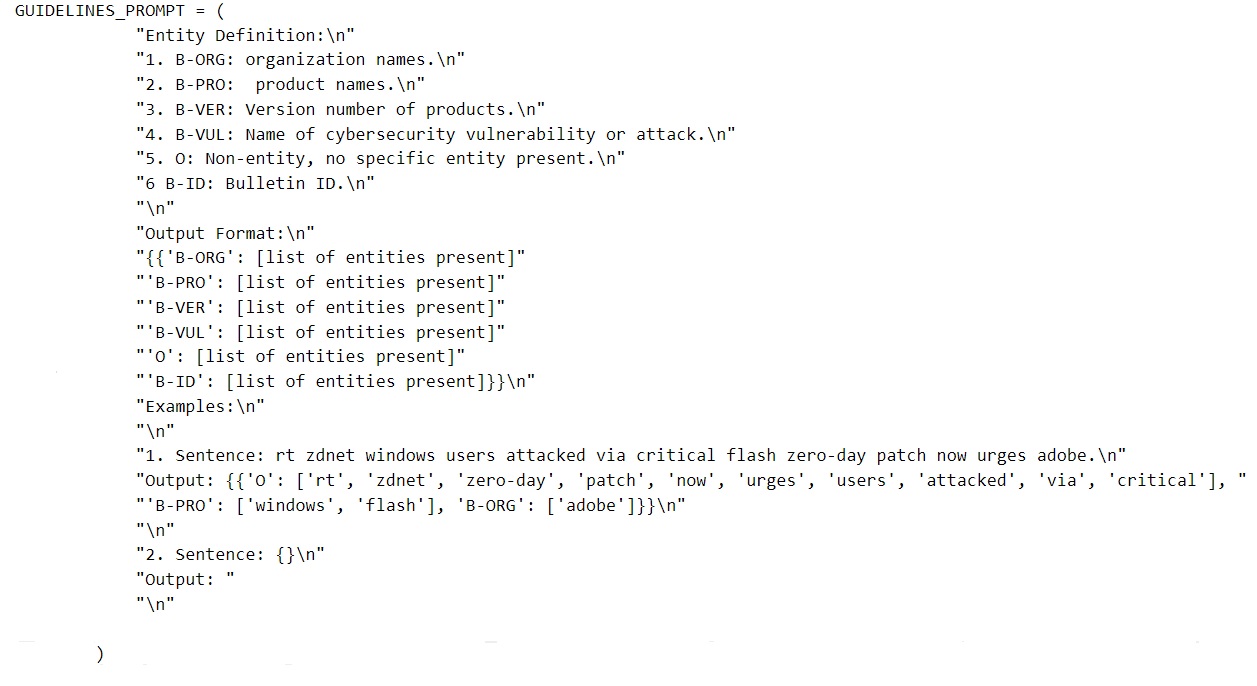}
    \caption{GLP NER approach: employing a ChatGPT-4 guideline prompt template}
    \label{fig:template}
\end{figure*}

The template for the guideline prompt used in the ChatGPT GLP NER approach is shown in Fig \ref{fig:template}.

\section{}
\label{AppendixB}
As described in Section~\ref{sec:data}, for each collected tweet a dataset entry is generated, including timestamp, keywords, original tweet, pre-processed tweet, cybersecurity relevance binary label, and sequence of named entities in the pre-processed tweet. Table \ref{tab:tweets} presents two examples of dataset entries.
In the relevance column, `1' denotes an entry considered relevant for cybersecurity, a `0' means otherwise. The last column shows the tags used to label the different NER entities.
\begin{table}[]
\centering
\caption{Samples from the 31281 tweet entries in the dataset}
\label{tab:tweets}
\begin{tblr}{
    colspec = {X[-1,l] l X[2,l] X[2,l] c X[l]},
  row{1} = {Alto,c},
  cell{2}{2} = {c},
  cell{2}{5} = {c},
  cell{2}{6} = {c},
  cell{3}{2} = {c},
  cell{3}{5} = {c},
  hline{1-4} = {-}{},
}
timestamp & keywords & original tweet & pre-processed tweet & relevance & entities\\
2018-07-24 01:00:46+00:00 & oracle & RT \@Oracle: Learn to use and understand \#Oracle’s Internet Intelligence Map https://t.co/l06Nyf1FFF \@Dyn https://t.co/uzozFKwm97 & rt oracle learn to use and understand oracle s internet intelligence map dyn & 0 & -\\
2016-12-09 19:19:38+00:00 & internet explorer & threatmeter: [dos] - Microsoft Internet Explorer 9 MSHTML - CDisp Node::Insert Sibling Node Use-After-Free (MS13-0... https://t.co/gLvEwpDL9v & threatmeter dos microsoft internet explorer 9 mshtml cdisp node::insert sibling node use-after-free ms13-0 & 1 & O O B-ORG B-PRO I-PRO B-VER O O O O O B-VUL B-ID
\end{tblr}
\end{table}

\section{}
\label{AppendixC}
The confusion matrices are provided the test and model combinations considered, excluding the 7B parameter models in Table~\ref{tab:accuracy} which achieved the worse results in the respective group. The rows in the matrices correspond to the actual expected result, whereas the columns show the predicted results.
\renewcommand{\thefigure}{C\arabic{figure}}    
\setcounter{figure}{0}  
\begin{figure*}[]
    \centering
    \includegraphics[width=\textwidth,keepaspectratio]{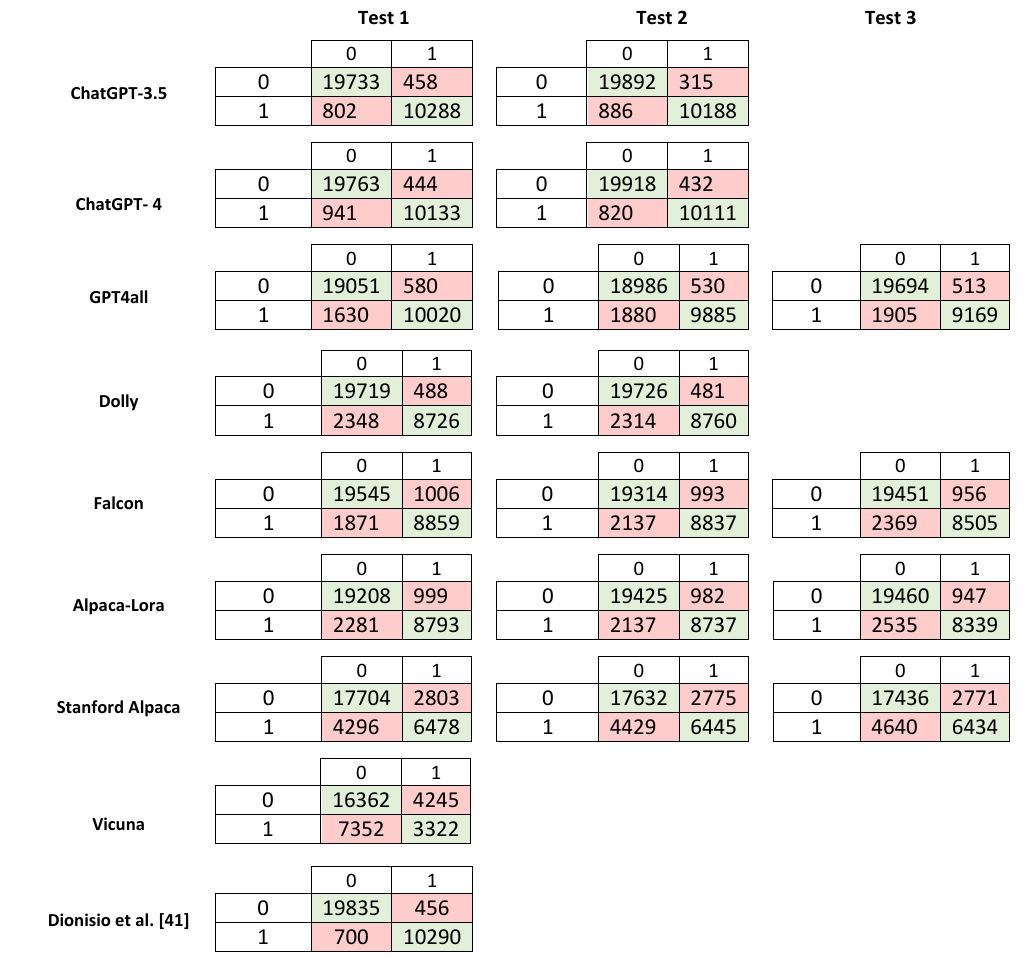}
    \caption{Confusion matrix of binary classification task}
    \label{fig:cm}
\end{figure*}
\end{appendices}
\end{document}